\documentclass[12pt]{article} 
\usepackage{epsfig}
\usepackage{amsfonts}
\usepackage{latexsym}
\usepackage{amsmath,amssymb}
\usepackage{verbatim}

\usepackage[textheight=9in, textwidth=6.5in, letterpaper]{geometry}

\numberwithin{equation}{section}

\def\ip{${\cal I}^+$}
\def\ipp{${\cal I}^+_+$}

\def\e{{\epsilon}}
\def\S{\Sigma}
 \def\p{\partial}
 \def\bz{{\bar z}}
 
 \def\bw{{\bar w}}
\def\[{\big[}
\def\]{\big]}
\def\o{\omega }

 \def\cL{{\cal L}}

\def\rz{{\rm z}}\def\rw{{\rm w}}
\def\>{|0\rangle }
\def\<{\langle 0|}
\newcommand{\bea}{\begin{eqnarray}}
\newcommand{\eea}{\end{eqnarray}}
\newcommand{\be}{\begin{equation}}
\newcommand{\ee}{\end{equation}}
\newcommand{\ba}{\begin{align}}
\newcommand{\ea}{\end{align}}

\renewcommand{\epsilon}{\varepsilon}

   \makeatletter
  \let\over=\@@over \let\overwithdelims=\@@overwithdelims
  \let\atop=\@@atop \let\atopwithdelims=\@@atopwithdelims
  \let\above=\@@above \let\abovewithdelims=\@@abovewithdelims
\renewcommand\section{\@startsection {section}{1}{\z@}%
                                   {-3.5ex \@plus -1ex \@minus -.2ex}
                                   {2.3ex \@plus.2ex}%
                                   {\normalfont\large\bfseries}}

\renewcommand\subsection{\@startsection{subsection}{2}{\z@}%
                                     {-3.25ex\@plus -1ex \@minus -.2ex}%
                                     {1.5ex \@plus .2ex}%
                                     {\normalfont\bfseries}}

\begin{document}
\begin{titlepage}
\unitlength = 1mm
\ \\

\begin{center}

{ \LARGE {\textsc{Asymptotic Symmetries of Yang-Mills Theory }}}

\vspace{0.8cm}
Andrew Strominger

\vspace{1cm}

{\it  Radcliffe Institute for Advanced Study, Harvard University,\\
Cambridge, MA 02138, USA}

\begin{abstract}
Asymptotic symmetries at future null infinity (\ip ) of Minkowski space for  electrodynamics with massless charged fields, as well as non-Abelian gauge theories with gauge group $G$, are considered at the semiclassical level. The possibility of charge/color flux through \ip\ suggests the symmetry group is infinite-dimensional. It is conjectured that the symmetries include a $G$ Kac-Moody symmetry whose generators are ''large" gauge transformations which approach locally holomorphic functions on the conformal two-sphere at \ip\ and are invariant under null translations. The Kac-Moody currents  are constructed from the gauge field at the future boundary of \ip. The current Ward identities include Weinberg's soft photon theorem and its colored extension.
\end{abstract}
\vspace{0.5cm}

\vspace{1.0cm}

\end{center}
\vspace{2.0cm}
\end{titlepage}

\pagestyle{empty}
\pagestyle{plain}

\def\vx{{\vec x}}
\def\ip{${\cal I}^+$}
\def\p{\partial}
\def\po{$\cal P_O$}
\def\0{{(0)}}
\def\1{{(1)}}
\def\2{{(2)}}

\pagenumbering{arabic}

\tableofcontents
\section{Introduction}

 Over half a century ago, Bondi,  van der Burg, Metzner and Sachs \cite{bms} argued that  gravitational wave scattering in asymptotically Minkowskian spacetimes is governed by an infinite-dimensional symmetry algebra now known as the BMS algebra. 
 The symmetry has consequences for any theory which can be coupled to gravity, and would be expected to play a central role in Minkowskian scattering theory. However, while some important results have been obtained, the full import of BMS symmetry has remained elusive. 
 
 The BMS generators are infinitesimal diffeomorphisms which preserve a prescribed structure at future null infinity, or \ip, but act nontrivially on the physical data residing there. \ip\ is the product of a  conformal two-sphere (directions of outgoing null rays)  with a null line (retarded times). The existence of infinitely many conformal two-spheres in \ip\ leads to both an infinity of charges - e.g. the Bondi masses for each $S^2$- and an infinite-dimensional symmetry group. The BMS algebra has two types of elements. First there are the global $SL(2,C)$ conformal transformations of the $S^2$ which are related to  Lorentz transformations. Second there are ''supertranslations" along the the null generators of \ip. These  are infinite in number because their magnitude can vary over the $S^2$. 
 
In a fascinating recent proposal, Barnich and Troessaert \cite{bt} and Banks \cite{banks} argued that the BMS algebra is, after all these years,  $not$ in fact the complete symmetry algebra of asymptotically Minkowskian spacetimes.  In the original analysis of \cite{bms}, it was required that all diffeomorphisms be everywhere finite on \ip.  In \cite{bt, banks} it was argued that this should be relaxed to allow for certain analytic singularities, or equivalently have holomorphic behavior on local patches of the conformal $S^2$. This relaxation enhances the $global$ $SL(2,C)$ subalgebra of BMS to two $local$ Virasoro algebras, leading to a new ``extended BMS symmetry''.  

Two decades after BMS, BPZ \cite{Belavin:1984vu} studied a mathematically similar problem: 2D field theories with a global $SL(2,C)$ symmetry. Unlike BMS, BPZ allowed symmetries with analytic singularities, which leads to two Virasoro algebras. The implications of this infinite symmetry algebra for 2D field theory are of course truly extraordinary. One hopes by analogy that the extended BMS symmetry has extraordinary implications for Minkowski scattering. It is plausible, as we explore in \cite{asip}, that some of these implications are already known in the guise of Weinberg's soft graviton theorem \cite{steve}. 

This paper will address a simpler,  but very closely related, toy version of the gravitational problem.  In electrodynamics with no massless charged particles, the symmetry group at both spatial infinity and \ip\ is a global $U(1)$. The $U(1)$ symmetry is generated by the associated conserved electric charge, which may be measured at either spatial infinity or \ip.

The situation is radically different if there are massless charged particles. Then there can be charge flux through \ip, and many different electric charges associated to the many asymptotic $S^2$s. Evidently, the \ip\ symmetry  is not a simple global $U(1)$. This raises our basic question:
\vskip.5cm
\noindent{\it What are the asymptotic symmetries at \ip\ of electrodynamics with massless charged particles?}
\vskip.5cm
\noindent This is a toy version of the question asked by BMS. It is of interest in its own right as well as for a warm-up to the gravitational  case. Part of this problem is defining both what is meant by asymptotic symmetries and how they act. 

We conjecture  herein that these symmetries include\footnote{ The conjecture does not preclude the possibility that the full asymptotic symmetry group is larger. In particular, one might consider, in analogy with the BMS supertranslations,  gauge parameters which locally approach arbitrary (not necessarily holomorphic) functions at infinity. Such transformations would not show up in the analysis of thus paper because they are ruled out by our gauge choice.} ''large" $U(1)$ gauge transformations which approach (the 
real part of) a holomorphic  function on local patches of the conformal $S^2$ on \ip\ and are constant along the null generators.  These are the toy analogs of the singular conformal transformations of \cite{bt}, and appear related to the Sky transformations of \cite{bal}.  A current $J_z$ tangent to the conformal $S^2$ at the future of \ip\ is  constructed in radiation gauge from the  asymptotic gauge field.   Insertions of contour integrals of this current are then shown to generate infinitesimal large gauge transformations on the matter fields and obey a $U(1)$ Kac-Moody algebra. Moreover the large gauge transformation law of the current itself hints at a nonzero Kac-Moody level proportional to ${1 \over  e^2}$.\footnote{With a current normalization fixed by the periodicity of the gauge transformations.}  While intriguing,  a precise statement about a possible level must await a better understanding of the putative CFT$_2$ at \ip: for example determining the sign (or phase) requires an as-yet-to-be-discussed notion of adjoint.  An asymptotic  Kac-Moody symmetry  fits well with the proposed extended BMS group, since  the generators are ready to transform under the Virasoro generators which appear when gravity is coupled.

Much of this structure is in fact already known in a different language. Around the same time as BMS, Weinberg \cite{steve} derived soft-photon theorems which give relations among scattering amplitudes with and without soft photons. In a key motivation for the present work, Maldacena and Zhiboedov have shown \cite{juan} by transforming Weinberg's momentum-space formulae into position space coordinates on \ip,  that the soft photon theorem can be rewritten in the form of a current algebra Ward identity. We will see that the relations among scattering amplitudes implied by  Weinberg's theorem can be viewed as a consequence of the large gauge symmetry. 

Weinberg does not encounter a non-zero level when considering multiple soft photon insertions. This is because his soft photon insertions are subtly different from our $J_z$:  they are gauge invariant but nonlocal while $J_z$ is local but not gauge invariant (see section 2.3). 

The situation for non-Abelian gauge theories (in the Coulomb phase) is similar. In general color flux can pass through \ip, 
leading to the basic question
\vskip.5cm
\noindent{\it What are the asymptotic symmetries at \ip\ for non-Abelian gauge theories?}
\vskip.5cm
\noindent The analysis is very similar to the abelian case with charged massless particles: for simple gauge group $G$ we find a
$G$ Kac-Moody algebra and again hints of a non zero level proportional to  ${1 \over g_{\rm YM}^2}$.

The analysis of this paper is restricted to the simplest nontrivial situations, as these already turn out to be quite subtle and rich. For example we work only at the semiclassical level and restrict to configurations which have no incoming charge flux on ${\cal I}^-$ and decay to the vacuum in the far future. 

This paper is organized as follows. Section 2 treats the $U(1)$ case. In 2.1 we describe the boundary conditions and falloffs at \ip\ in radiation gauge, and develop the asymptotic expansion in the inverse radius.  Attention is drawn to residual large gauge transformations which are locallly holomorphic functions on $S^2$.  2.2 briefly describes the conformal transformation properties of the asymptotic data.  In 2.3 we define  the Kac-Moody current $J_z$ in terms of tangential components of the gauge field at the infinite future of \ip, denoted \ipp. $J_z$ is shown semiclassically to acquire poles where charged particles cross \ip, as appropriate to such a current. In 2.4 we recast the appearance of poles in $J_z$ as a semiclassical Ward identity and discuss the relation to Weinberg's soft photon theorem. It is further shown that insertion of $J_z$ contour integrals generate large gauge transformations of matter fields on local patches of the $S^2$ at \ip. 2.5 reviews the radiation-gauge Feynman two-point function for gauge field in our notation and computes a boundary gauge field correlator. 2.6 infers a nonzero Kac-Moody level from the coefficient of the gauge variation of $J_z$ itself.  The nonabelian case is very similar and sketched in section 3. Finally in section 4 we discuss the realization of this structure within string theory and speculate that, in the spirit of  \cite{Giveon:1998ns}, the spacetime Kac-Moody current $J_z$ is the lift of the familiar worldsheet Kac-Moody current associated to spacetime gauge symmetries. 

\section{Electrodynamics}
In this section we consider a $U(1)$ gauge theory coupled to massless charged matter. 
The Maxwell electromagnetic action is
\be \label{max}S_M=-{1 \over 4e^2}\int d^4x\sqrt{-g}F_{\mu\nu}F^{\mu\nu}.\ee
This has the gauge symmetry
\be \delta_{\hat \e} A_\mu=\p_\mu {\hat \e}\ee
 with the periodicity \be \label{pd} {\hat \e}\sim {\hat \e}+2\pi.\ee
Given a surface $\S$ in \ip, electromagnetic Bondi charges can be defined by\footnote{$Q^M$ is an integer with the given $\hat \e$-periodicity. $Q^E$ differs by a factor of $e$ from some definitions of the electric charge, and is also set to an integer by Dirac quantization.}
\be Q^E={1 \over e^2}\int_\S * F,~~~~~~~Q^M={1 \over 2\pi}\int_\S F.\ee
We consider the case $Q^M=0$ for all $\Sigma$.
In a theory with massless electrically charged particles which reach \ip,  $Q^E$ will depend on the choice of $\S$.
We will work with the flat Minkowski metric in coordinates 
\be \label{flt}ds_F^2=-dv^2-2drdv+2r^2\gamma_{z\bz}dzd\bz,~~~~\int d^2z\gamma_{z\bz}=4\pi. \ee
 $(v, z,\bz)$ with $v= t-r$, are coordinates on \ip. For the special case $z=e^{i\phi}\tan {\theta \over 2}, ~~\gamma_{z\bz}=2(1+z\bz)^{-2}$.  The nonzero connection coefficients are 
 \be \Gamma^z_{rz}={1\over r},~~~ \Gamma^z_{zz}=\p_z\ln\gamma_{z\bz}     , ~~~\Gamma^v_{z\bz}={r\gamma_{z\bz}} , ~~~\Gamma^r_{z\bz}=-{r\gamma_{z\bz}}.
 \ee

Denoting the charges and surfaces at retarded time $v$ by $Q^E(v)$ and $\S_v$ 
one has
\be Q^E(v)={1 \over e^2}\int_{\S_v} d^2z \gamma_{z\bz}r^2F_{rv}.\ee
For nonzero matter charge  current  $j_\nu^M$  the Maxwell equations $\nabla^\mu F_{\mu\nu}=e^2j_\nu^M$ are
\be \label{mao} -\gamma_{z\bz}r^2\p_v F_{rv}+\p_zF_{\bz v}+\p_\bz F_{zv}+\p_r(\gamma_{z\bz}r^2 F_{rv})=e^2\gamma_{z\bz}r^2j_v^M,\ee
\be \label{mabo} \p_zF_{\bz r}+\p_\bz F_{zr}+\p_r(\gamma_{z\bz}r^2 F_{rv})=e^2\gamma_{z\bz}r^2j_r^M,\ee
\be \label{maco} r^2\p_r( F_{rz}-F_{vz})-r^2\p_vF_{rz}+\p_z\big(\gamma^{z\bz}F_{\bz z}\big)=e^2 r^2j_z^M.\ee
We then have
\be \p_v Q^E(v)= - \int_{\S_v} d^2z \gamma_{z\bz}r^2j^M_{v}.\ee
The causal geometry is depicted in figure 1. 

In this paper we impose a number of simplifying restrictions in order to exhibit the basic phenomena in the simplest possible setting. 
We consider only theories which have no stable massive particles and revert to the vacuum in the far future. In the spirit of \cite{bms} we concentrate  on the structure of \ip\ and ignore the parallel structure on ${\cal I}^-$, as well as the scattering between the two.  We specialize to the case where there is no incoming charges, namely on ${\cal I}^-$
\be \label{rs} j^{M}_\mu(r,-\infty, z,\bz)=0,  ~~~~~Q^E(-\infty)=0. \ee
At the quantum level, this means the incoming state is annihilated by the local charge current. 
This is of course semiclassically consistent with nontrivial outgoing charge fluxes on \ip, subject to the restriction that the total integrated charge flux must vanish because the system reverts to the vacuum. 
We expect it is both possible and interesting to extend our analysis to more general cases.
\begin{figure}[t!]
\centering{
\includegraphics[width=200mm,height=160mm]{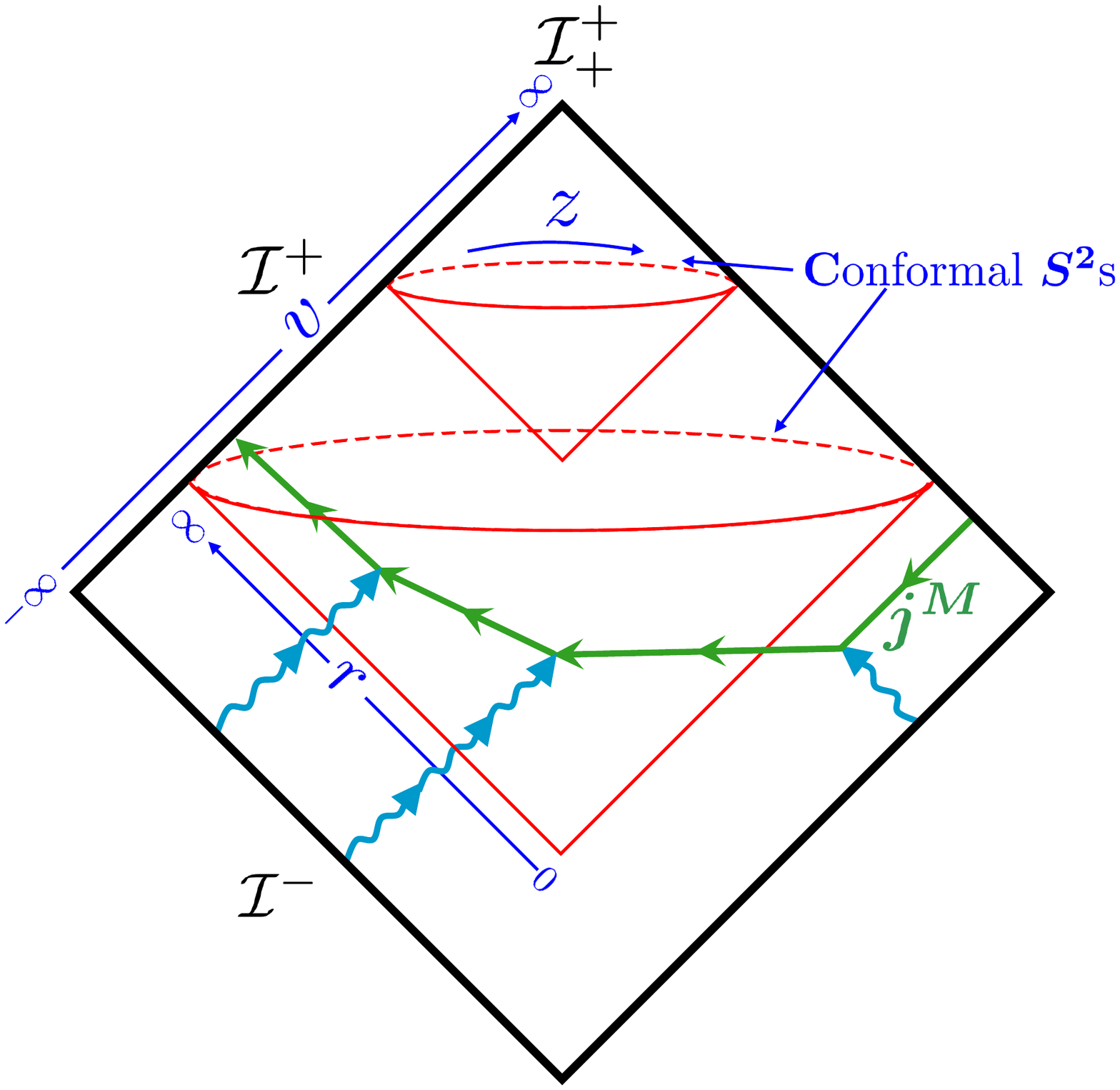}}
\caption{Penrose diagram for Minkowski scattering. The figure depicts incoming photons (blue) from ${\cal I}^-$ ($v=-\infty$) creating a matter charge current $j^M$ (green) outgoing at \ip ($r=\infty$). The total charge $Q^E(v)$ is measured on a conformal sphere at \ip  (parameterized by $(z,\bz)$),  and depends on the  retarded time $v$.  The (red) cones are null surfaces of constant $v$.}
\label{fig1}
\end{figure}

\subsection{The boundary data problem}
In this subsection we consider the inverse of the initial data problem on ${\cal I}^-$, namely the final data problem on \ip.  It is convenient to impose the temporal gauge condition 
\be \label{rg} A_v=0,\ee which becomes $A_t$=0 in standard Minkowskian coordinates.
This leaves unfixed a residual $v$-independent gauge symmetry. We partially fix this by demanding in the far past :\footnote{This condition is obstructed when there is incoming charge flux on ${\cal I}^-$.}
\be \label{xdc}r^2\gamma_{z\bz} \nabla^\mu A_\mu(z,\bz,r,-\infty)=\big [ \p_z A_\bz+\p_\bz A_z+(\p_r-\p_v)(\gamma_{z\bz}r^2 A_r)\big]_{{\cal I}^-}=0,\ee    (\ref{rg})-(\ref{xdc}) is commonly referred to as radiation gauge. Still remaining 
are gauge transformations obeying 
\be \label{sxd}2 \p_z\p_\bz{\hat \e}+\gamma_{z\bz}\p_r( r^2 \p_r{\hat \e})=0=\p_v{\hat \e}.\ee
Locally this allows the residual transformations
\bea\label{zs}{\hat \e}&=&\e(z)+\bar \e(\bz),\cr
\delta_\e A_z&=&\p_z\e.\eea 
for holomorphic $\e$.
Globally of course there are no holomorphic functions on $S^2$ except the constant, and (\ref{zs}) must have singularities at points. 
Nevertheless we shall consider below transformations in a patch on \ip\ for which the local solutions (\ref{zs}) are relevant. The transformations (\ref{zs}) are the electromagnetic analog of the local  conformal transformations proposed in \cite{bt} as part of the extended BMS group.
They create zero-momentum  "boundary photons" and we will refer to them as large gauge transformations.

In radiation  gauge one has  the simple relations
\be F_{vz}=\p_vA_z, ~~~~F_{vr}=\p_vA_r .\ee The Maxwell equations may then be written
\bea \label{maa} -\p_v[\p_zA_\bz+\p_\bz A_{z}+(\p_r-\p_v)(\gamma_{z\bz}r^2 A_r)]=e^2\gamma_{z\bz}r^2j_v^M,\eea
\bea \label{mab}2\p_z\p_\bz A_r-\p_v \p_r(\gamma_{z\bz}r^2 A_{r})-\p_r(\p_zA_\bz+\p_\bz A_{z}) =e^2\gamma_{z\bz}r^2j_r^M,\eea
\bea \label{mac}r^2(\p_r^2 A_z-2\p_r\p_vA_z)-r^2\p_r\p_zA_r+r^2\p_v\p_zA_r+\p_z\big(\gamma^{z\bz}(\p_\bz A_z-\p_z A_\bz)\big) =e^2r^2j_z^M.\eea

A typical wave solution near \ip\ is of the general form
\be A_r=0,~~A_z=a(z,\bz)e^{ikv}.\ee
This motivates the asymptotic boundary condition at \ip
\be A_z\sim O(1),~~~A_r \sim O({1 \over r^2}). \ee 
This  leading behavior of $A_r$ insures that $Q^E(v)$ is always finite.
We may then expand near \ip
\bea A_z(r,v,z,\bz)&=&\sum_{n=0}^\infty {A^{(n)}_z(v,z,\bz)\over r^n},\cr
 A_r(r,v,z,\bz)&=&\sum_{n=0}^\infty {A^{(n)}_r(v,z,\bz)\over r^{n+2}}.
 \eea
 For simplicity we consider the case where $j_r^M=j_z^M=0$ near \ip. Current conservation then implies in this region \be  j^M_v(r,v,z,\bz)= {j^{M(0)}_v(v,z,\bz)\over r^2}.\ee  
 The leading terms at large $r$ in (\ref{maa})-(\ref{mac}) then give the equations on \ip
 \be\label{ko} -\p_v\big[\gamma^{z\bz} (\p_\bz A^\0_z +\p_zA^\0_\bz)-\p_vA^\0_r \big]= e^2 j^{M\0}_{v},  \ee 
\be 2\p_vA^\1_z+\p_v\p_zA_r^\0+\p_z\big(\gamma^{z\bz}(\p_\bz A^\0_z -\p_zA^\0_\bz)\big)=0\ee
\be \label{madb}2\p_z\p_\bz A_r^\0+ \gamma_{z\bz}\p_v A_{r}^\1+\p_zA^\1_\bz+\p_\bz A^\1_{z}=0\ee
The boundary fields appearing here are  constrained by the subleading term in (\ref{maa})
\be\label{dsaz} -\p_v\big[\gamma^{z\bz} (\p_\bz A^\1_z +\p_zA^\1_\bz)-\p_vA^\1_r \big]=0.\ee
Integrating the constraint equation  (\ref{ko}) along a null generator of \ip\ and using (\ref{xdc}) gives 
the important relation \be\label{kop} 2\p_\bz A^\0_z = -\gamma_{z\bz} e^2\int_{-\infty}^v dv' j^{M\0}_{v'}+\gamma_{z\bz} F^\0_{vr}-F^\0_{z\bz}. \ee 
Integrating this once more with respect to $v$  gives  
$A_r^\0$ in terms of $A^\0_z$. The other equations of motion determine 
$A^\1_z$ in terms of $A^\0_z$
 \be\label{sx} A^\1_z=-\p_z\int_{-\infty}^v dv'\Big[\gamma^{z\bz}\p_\bz A^\0_z+  {e^2\over 2}\int_{-\infty}^{v'} dv'' j^{M\0}_{v''}
\Big]\ee
Finally integrating (\ref{dsaz}) with respect to $v$  then gives  
$A_r^\1$ in terms of $A^\0_z$. Further subleading  equations of motion similarly determine further subleading terms in the power expansion. 

Hence the local free final data lies in the two unconstrained real functions $A_z^\0$.  This is as expected, because a massless spin one has two degrees of freedom and should be locally specified by two real (or one complex) functions on \ip. Noting that 
 $\p_vA_z^\0=F_{vz}^\0$, the free data is simply related to the radiation energy flux at \ip\  according to $T_{vv} \sim {1 \over r^2}\gamma^{z\bz}F^\0_{vz}F^\0_{v\bz}$.  
 \subsection{Conformal transformations}

  In this subsection we briefly describe the action of \ip\ conformal transformations  - a subset of the extended BMS transformations - on fields appearing in the ${1\over r}$ expansion around \ip.  Similarities with the familiar AdS case will be evident. The transformations are generated by the real part of the complex vector fields \cite{bt}
  \bea \label{cnt} \zeta^a_{\rm conf}\p_a &\equiv&(\zeta^z-{v \over 2r} \gamma^{z\bz} D_\bz D_z\zeta^z)\p_z+(-{r\over 2}D_z\zeta^z+{v\over 4}D^2 D_z\zeta^z)\p_r -{v\over 2 r}\gamma^{z\bz} D^2_z\zeta^z\p_\bz+{v\over 2 }D_z\zeta^z\p_v \cr
  &=&(1+{v R \over 4 r} )\zeta^z\p_z-({r \over2}D_z\zeta^z+{v \over 4 } {D_z(R\zeta^z)})\p_r -{v\over 2 r}\gamma^{z\bz} D^2_z\zeta^z\p_\bz+{v\over 2 }D_z\zeta^z\p_v \cr  &=&(1+{v  \over 2r} )\zeta^z\p_z-(1+{v  \over  r} ){D_z\zeta^zr\over 2}\p_r -{v\over 2 r}\gamma^{z\bz} D^2_z\zeta^z\p_\bz+{v\over 2 }D_z\zeta^z\p_v ,\eea
  where $\p_\bz\zeta^z=0$, $D$ is the $\gamma$-covariant derivative with curvature $R$ and $D^2=\gamma^{z\bz}(D_zD_\bz+D_\bz D_z)$.  In the last line we use the fact that $R=2$ for our choice of $\gamma_{z\bz}$. These act on the flat Minkowski metric as
  \be \cL_\zeta ds^2=-rvD^3_z\zeta^zdz^2,\ee
which is a subleading (in $ 1 \over r$) contribution to the induced $S^2$ metric.
 A massless scalar field has the expansion near \ip\ 
 \be \phi(r,v,z,\bz)=\sum_n{\phi^{(n)}(v,z,\bz)\over r^{n+1}}.\ee
 Conformal transformations act on the leading expansion coefficient as 
 \be \cL_\zeta \phi^{(0)}(v,z,\bz)={1 \over 2}D_z\zeta^z (1+v\p_v)\phi^{(0)}(v,z,\bz)+\zeta^z\p_z \phi^{(0)}(v,z,\bz).\ee
 Hence the $v$ zero modes of these coefficients  transform as two dimensional conformal fields with  with weights $(h,\bar h)=({1\over 2},{1 \over 2})$.
For the vector field one finds the $v$ component obeys
\be \cL_\zeta A_v=\zeta^a_{\rm conf}\p_a A_v+{1\over 2r} (\zeta^zA^\0_z-\gamma^{z\bz} D^2_z\zeta^zA^\0_\bz)+\cdots \ee
The gauge choice $A_v=0$  employed in this paper is not preserved by ${\cal L}_\zeta$ and explicitly breaks the symmetry. Hence, while useful for other purposes, it is not ideal for analyzing conformal properties. More generally the extended BMS symmetry potentially  constrains the  boundary correlations functions on \ip\ and the S-matrix.  Indeed, the peeling theorems of Newman and Penrose \cite{np}  exploit the original BMS symmetry to constrain the behavior of gravitational radiation  at \ip.
\subsection{The boundary current}
We consider final data which corresponds to solutions with finite total energy as measured at spatial infinity as well as vanishing initial and final charges. This requires that at the future ($v\to \infty$) of \ip (denoted \ipp\ and depicted in figure 1)  the field strengths go to zero: 
\be F^\0_{z\bz}(\infty,z,\bz)=0=F^\0_{rv}(\infty,z,\bz),\ee
or equivalently 
\be \p_z A^\0_\bz(\infty,z,\bz)=\p_\bz A^\0_z(\infty,z,\bz),~~~~~\p_vA^\0_r(\infty,z,\bz)=0.\ee
(\ref{kop}) then reduces to
\be \p_\bz A^\0_z(\infty,z,\bz)  =-{\gamma_{z\bz}{e^2\over 2}}\int_{-\infty}^\infty dv' j^{M(0)}_{v'}.\ee
For such finite-energy data we can define currents on \ipp\  
\be \label{dsf} J_z(z,\bz)=-{4\pi \over e^2}A^\0_z(\infty,z,\bz).
\ee
A parallel construction is possible for $J_\bz$.
This current is sourced by charges which cross \ip\ at any value of retarded time
\be \p_\bz J_z(z,\bz)={2\pi } \gamma_{z\bz}\int_{-\infty}^\infty j^{M\0}_{v'}(v',z,\bz)dv'. \ee

Consider for  example of a set of massless point electric particles crossing  \ip\ with the conserved charge current 
\be r^2j_v^M=   \sum_kq_k\delta(v-v_k){\delta^2(z-z_k)\over \gamma_{z\bz}} , \ee
where \be \sum_k q_k=0,\ee
so that  the total initial and final electric charge vanishes. 
We then have \be \p_\bz J_z= {2\pi }\sum_k q_k
\delta^2(z-z_k).\ee
Using $\p_\bz{1 \over z}=2\pi \delta^2(z)$ this is solved by 
\be  \label{jw}J_z=\sum_k { q_k\over z-z_k}.\ee
Locally, a holomorphic function can be added corresponding to large gauge transformations. 
We will see in the next subsection that the quantum version of this simple equality is the Ward identity for a $U(1)$ current algebra
on \ipp. 

Note that in radiation gauge the presence of pointlike charged radiation passing through \ip\ implies the existence of singularities in the gauge potential $A^\0_z\to -{e^2 \over 4\pi}J_z$ even at  \ipp\ where the system reverts to the vacuum.  Smoothing out the charge flux over $z$ will also smooth out $A^\0_z$, but it will necessarily  remain nonzero in the far future.

It is interesting to compare the current $J_z$, viewed as a quantum operator,  to the  zero momentum limit of an operator annihilating a photon emerging at $z$ on \ip.   
The gauge invariant operator \be\label{sf} V_{z,\omega }(z,\bz)= \int_{-\infty}^\infty dv e^{-i\omega v}F^\0_{vz}(v, z, \bz),~~~\omega >0\ee  annihilates a positive helicity photon of energy $\o$ emerging at $z$ while the Fourier transform of  $F^\0_{v\bz}$  annihilates negative helicity photons. In the soft photon limit $\o\to 0$  (\ref{sf}) becomes a total derivative and reduces to  \be \label{gi}V_{z,0 }(z,\bz)= A^\0_{z}(\infty, z, \bz)-A^\0_{z}(-\infty, z, \bz). \ee
This is closely related  to the current $J_z$ defined in (\ref{dsf}). Given our boundary conditions including (\ref{rs}) both are sourced by charge fields according to (\ref{jw}).
However there is a difference: $V_{z,0 }$ is nonlocal on \ip\ but invariant under large gauge transformations, while $J_z$ is local but transforms nontrivially under large gauge transformations.

\subsection{Ward identities and Weinberg's theorem}
In this subsection we recast the pole formula 
(\ref{jw}) for the current as a semiclassical relation between correlation functions, identify it as a Kac-Moody Ward identity and relate it to Weinberg's soft photon theorem.

S-matrix elements with no charged particles coming in from ${\cal I}^-$ and $n$ charged particles with charges $q_k$ going out at $(v_k, z_k,\bz_k)$  on \ip\ are constructed  from  correlation functions of the form 
\be \< V_1(v_1,z_1,\bz_1)...V_n(v_n, z_n,\bz_n)....\> .\ee
Here $V_k$ is an operator at $r=\infty$ on \ip\ which annihilates a matter particle of charge $q_k$ at $(v_k,z_k,\bz_k)$.\footnote{ It would be interesting to work out in detail the S-matrix in a basis with particles labelled by their $(z,\bz)$ coordinates on ${\cal I}^\pm$ (rather than the usual momentum space labels)  and study their conformal transformation properties. This is partially acheived in the twistor formulation.} As the counterpart of our assumption (\ref{rs}) of no charge flux across ${\cal I}^-$, the incoming particles, denoted by $....$,  are all taken here to be neutral.
Arbitrary time and radius independent  gauge transformations $f(z,\bz)$ of the matter fields on any fixed time slice with $v+r=t_0$ are generated by the matter part of Noether charge 
\be Q_{t_0}^M[f]=\int_{v+r=t_0} d^3\Sigma^\mu f j^M_\mu \ee
according to 
\be \[Q_{t_0}^M[f], V_k(r,t_0-r,z,\bz)\]=-f(z,\bz)q_kV_k(r,t_0-r,z,\bz).\ee
Consider  
the normal-ordered $v$-zero mode of the matter charge current on \ip  
\be q_{z\bz}(z,\bz)=\gamma_{z\bz}\int_{-\infty}^\infty  dv j^{0M}_v(v,z,\bz)
. \ee Since the fixed time surface approaches \ip\ in the limit $t_0\to \infty$, 
$q_{z\bz}$ is the special case of $Q^M_\infty[f]$ where $f$ is set to a delta function. It follows that \be [q_{z\bz}(z,\bz),V_k(v_k,z_k,\bz_k)]=-q_kV_k(v_k,z_k,\bz_k)\delta^2(z-z_k).\ee
Using this and $\<q_{z\bz}=0$, insertions of 
$q_{z\bz}$ in correlation functions gives 
\be \< V_1(v_1,z_1,\bz_1)...V_n(v_n, z_n,\bz_n)q_{z\bz}(z,\bz)....\> =\sum_{k=1}^n  q_k \delta^2(z-z_k)  \< V_1(v_1,z_1,\bz_1)...V_n(v_n, z_n,\bz_n)....\> .\ee 
The integrated constraint (\ref{kop}) can be written 
\be \label{xs}\p_\bz J_z=2\pi q_{z\bz} +{2\pi \over e^2}F^\0_{z\bz}(\infty, z, \bz)-{2\pi \over e^2}\gamma_{z\bz}F^\0_{vr}(\infty, z, \bz). \ee
The last two terms are the magnetic and electric field operators in the infinite future at \ipp. If there are no operator insertions at \ipp, 
and the incoming state has finite extent and energy, these terms vanishes in vacuum correlation  functions and can be dropped.\footnote{These terms are relevant for multiple $J_z$ insertions on \ip. }
We thereby reproduce  the operator version of the semiclassical relation (\ref{jw})
\be\label{xzc} \< V_1(v_1,z_1,\bz_1)...V_n(v_n, z_n,\bz_n)J_z(z,\bz)....\> =\sum_{k=1}^n{q_k \over z-z_k} \< V_1(v_1,z_1,\bz_1)...V_n(v_n, z_n,\bz_n)....\> .\ee 
This is not a new result: Weinberg's soft photon theorem \cite{steve} for amplitudes reads
\be\label{xzfc} \< V_1(p_1)...V_n(p_n)V_\gamma(\e,p_\gamma))....\>_{LSZ} \sim \sum_{k=1}^n{q_k\e\cdot p_k\over p_\gamma\cdot p_k} \< V_1(p_1)...V_n(p_n)....\>_{LSZ}, ~~~~p_\gamma \to 0, \ee 
where the subscript denotes that correlators are LSZ reduced to amplitudes. 
Here the $n$ particles are in momentum eigenstates $p_k$ with $p_k^2=0$ and 
$V_\gamma$ annihilates a soft photon with momentum $p_\gamma$ and polarization $\epsilon$. 
Maldacena and Zhiboedov have shown 
\cite{juan} that, for an appropriate normalization of $V_\gamma$, (\ref{xzc}) is simply a rewriting of (\ref{xzfc}) in position space for the conformal sphere at \ip.  References \cite{steve} and \cite{juan} actually consider an insertion of the closely related gauge invariant zero-energy photon vertex operator, denoted $V_{z,0}$ in equation (\ref{gi}). This differs from $J_z$ by the insertion of $A_z$ at spatial infinity which will not give charge-dependent terms in  (\ref{xzc}) (but is relevant for multiple current insertions).

It is interesting to see how the collinear momentum space singularity in (\ref{xzfc}) becomes the pole in (\ref{xzc}) in the simplest context: a more detailed and complete treatment is in \cite{juan}. 
For this we need the transformation from our retarded coordinates to the usual flat Minkowski coordinates \bea \label{fm}ds^2&=&-dt^2+d\vec x \cdot d\vec x,\cr
t&=&v+r,\cr
x^1+ix^2&=&r{2z\over 1+z\bz},\cr
x^3&=&r{1-z\bz\over 1+z\bz},\eea
with $\vec x=(x^1,x^2,x^3)$ obeying  $\vec x \cdot \vec x =r^2$. 
At late times and large $r$  the wave packet for a massless particle with charge $q$ and spatial momentum centered around $\vec p$
becomes localized on the conformal  sphere near  the point 
\be {\vec p \over \omega}={\vec x \over r}={1 \over 1+z\bz}(z+\bz,-iz+i\bz,1-z\bz). \ee
where $\vec p\cdot\vec p=\omega^2$. Consider the example of a soft photon with momentum $p_\gamma^3=\omega_\gamma$
which hits the asymptotic sphere at $z=0$.
The positive helicity spatial polarization vector obeying $\vec \e \cdot \vec p_\gamma=0$ 
is
\be \vec \e=(1,-i,0 ).\ee
These obey
\be p_\gamma \cdot p= -2\omega \omega_\gamma{z\bz\over (1+z\bz)}, ~~~~\e \cdot p= 2\omega {\bz\over (1+z\bz)}.\ee
The Weinberg soft factor is
\be {q \e\cdot p \over p_\gamma \cdot p}=-{1 \over \omega_\gamma}{q \over z},\ee
which exhibits the collinear pole at $z=0$ as in  (\ref{xzc}). The factor of $\omega_\gamma$ is ultimately cancelled in (\ref{xzc}) because it involves a zero mode of $F_{vz}$ rather than $A_z$ \cite{juan}. 

Let us now return to  (\ref{xzc}) and consider an integral of the current $J_z$ convoluted with $\e$ around a contour $C$ encircling all the operator insertions at  $(z_1,... z_n)$. One immediately finds. 
\bea\label{xzzc} \< V_1(v_1,z_1,\bz_1)...V_n(v_n, z_n,\bz_n){1 \over 2\pi i}\oint_C \e(z) J_zdz....\> &=&\sum_{k=1}^n q_k \e(z_k) \< V_1(v_1,z_1,\bz_1)...V_n(v_n, z_n,\bz_n)....\>\cr & =&-i\delta_\e  \< V_1(v_1,z_1,\bz_1)...V_n(v_n, z_n,\bz_n)\>.\eea 
Hence large gauge transformations in the region bounded by a contour $C$ are generated by insertions of  contour integrals $\oint_C J$. This identifies Weinberg's soft-photon relations between S-matrix elements as a Ward identity of the large gauge symmetry (\ref{zs}). 

\subsection{Green functions}
In this section we record the radiation-gauge two-point functions in the free quantum theory, and take their limits to obtain boundary correlators on \ip. The standard radiation  gauge $A_t=0=\nabla^\mu A_\mu$ is the same as $A_v=0=\nabla^\mu A_\mu$.
The  time-ordered Feyman two-point function is, in the flat Minkowski coordinates (\ref{fm})\footnote{See $e.g$ equation 14.54 in  \cite{bd}.}\be \label{zvb}{1 \over e^2}\<T(A_i(x)A_j(x'))\>= \delta_{ij}D_F(x,x')- \p_i\p'_j\Delta(x,x'),\ee
where $i,j=1,2,3$, $x\sim(t,\vec x)$,
\be D_F(x,x')=i\int {d^4 k \over (2\pi)^4}{e^{ik\cdot(x-x')}\over (k_0^2 -\vec k^2+i\e)}, ~~~\nabla^2D_F(x,x')=i\delta^4(x,x'), \ee
is the Feynman two-point function  for a massless scalar  and
\be \label{dsm} \Delta=-\vec \p^{-2}D_F,\ee
where
\bea \vec  \p^{-2}\delta^3(\vec x, \vec x')&=&-{1 \over 4 \pi \sqrt{(\vec x-\vec x')^2} }\cr &\equiv&-{1 \over 4 \pi \ell}.\eea
is the inverse of the spatial laplacian $\vec \p^2$. One finds 
\be \p_t^2 \Delta  =  {1\over 4\pi^2((|t-t'|+i\e)^2-\ell^2)}-{i\delta(t-t')\over 4 \pi \ell},\ee
Integrating twice and suppressing the $i\e$ yields 
\be \label{dlt}\Delta = {1\over 8\pi^2}\big({t-t'\over \ell} \ln {\ell -t+t'\over \ell+ t-t'}-\ln \big( \ell^2-(t-t')^2 \big)\big)-i{|t-t'|\over 8\pi \ell}.\ee
We have fixed the integration constant using (\ref{dsm}). 
In Bondi coordinates we have
\bea \ell^2&=&(r-r')^2+4rr'{(z-z')(\bz-\bz')\over (1+z\bz)(1+z'\bz')}.\eea 
Commutators of gauge-variant operators are in general nonlocal. For example it follows from (\ref{zvb}) that at equal times
\be\label{ccr}  [\p_tA_i(t,\vec x),A_j(t,\vec x')]=-ie^2\delta_{ij} \delta^3(\vec x,\vec x')-i\p_i\p'_j{ e^2 \over 4\pi \ell}.\ee

The boundary correlators are limits of  the two point functions  in which we take $r=r'\to \infty$ with the other coordinates fixed.  To leading order in $1 \over r$  \bea 
       \ell^2(x,x')&=&4r^2{ (z-z')(\bz-\bz')\over (1+z'\bz ')(1+z\bz)}+\cdots,\cr
       \Delta(x,x')&=&-{1 \over 8\pi^2}\big( 2\ln r+\ln { 4(z-z')(\bz-\bz')\over (1+z'\bz ')(1+z\bz)}+\cdots\big).\eea 
We then have
 \be \<A^\0_z(v,z,\bz)A^\0_z(v',z',\bz')\>_{{\cal I}^+}=-e^2 {\p^2 \Delta \over \p z \p z'}+e^2 \p_z \vec x \cdot \p_{z'}\vec xD_F(x,x')= {e^2 \over 8\pi^2 (z-z')^2}(1-{ \ell^2 \over 4r^2}).\ee

\subsection{Kac-Moody level?}

The Kac-Moody current $J_z$, which is proportional to the gauge potential, transforms non-trivially under large gauge transformations:\footnote{
In general the normalization of a $U(1)$ current is arbitrary. We have fixed this ambiguity by requiring equation (\ref{zs}) hold with the factor written and periodicity (\ref{pd}). }
\be \label{gt} \delta_\e J_z=- {4\pi \over e^2} \p_z\e.\ee
Often such a transformation law for a Kac-Moody current indicates a nonzero level, in this case proportional to 
${4\pi \over e^2}$.  However such an interpretation is too hasty here as we have not succeeded in obtaining this transformation law as some kind of commutator of $J_z$ with itself. Moreover the level can be thought of as the norm of identity descendants, and a norm has not been introduced yet. In the usual 4D field theory adjoint $J_z^\dagger =J_\bz$, which is not appropriate for a CFT$_2$ current.  These issues presumably must be resolved before inferring a level from (\ref{gt}).  Hence at this point we simply regard (\ref{gt}) as an interesting hint to guide further investigations.

 \section{The non-Abelian case}
The nonabelian case with group $G$ is very similar to the abelian one, except one must keep track of a few matrix orderings. 
We will only sketch the results. 
Radiation gauge for the nonabelian theory is defined by 
\be \label{rag}A_v=0,\ee
\be\label{ioa}\big[ \sqrt{-g}\nabla^\mu A_\mu\big]_{{\cal I}^-}=\big [ \p_z A_\bz+\p_\bz A_z-r^2\p_vA_r+\p_rr^2A_r\big]_{{\cal I}^-}=0.\ee 
This is the same as the Abelian case except that $A_\mu$ is now valued in the Lie algebra of $G$. We also assume here no color flux incoming on ${\cal I}^-$. 

Expanding near \ip
\bea A_z(r,v,z,\bz)&=&\sum_{n=0}^\infty {A^{(n)}_z(v,z,\bz)\over r^n},\cr
 A_r(r,v,z,\bz)&=&\sum_{n=0}^\infty {A^{(n)}_r(v,z,\bz)\over r^{n+2}}
,
 \eea 
 at leading order we obtain 
 \bea F^{\0}_{z\bz}&=&\p_zA^{\0}_\bz-\p_\bz A^{\0}_z +[A^{\0}_z,A^{\0}_\bz],\cr
  F^{\0}_{vz}&=&\p_v A_z^{\0},\cr
   F^{\0}_{rz}&=&-D^\0_zA_r^\0-A_z^{\1},\cr
  F^{\0}_{rv}&=&-\p_vA_r^\0,\eea
where $D^\0_z=\p_z+[A^\0_z, ~~]$.
Radiation gauge locally leaves unfixed large nonabelian gauge transformations  $\e(z)$:
\be\delta A^\0_z= D^\0_z\e.\ee
The constraint equation on \ip\ is
 \be\label{wxt} -\p_v\big(\p_\bz A^{\0}_z +\p_zA^{\0}_\bz-\gamma_{z\bz}\p_vA_r^{\0}\big)=[A^{\0}_z,\overleftrightarrow {\p_v}A^{\0}_\bz]+\gamma_{z\bz} g_{\rm YM}^2j^{\0 M}_v.  \ee 
 where $j^{0 M}_v$ is the leading component of any matter color current on \ip\ and the commutator is the gluon color current. Assuming the field strengths vanish at \ipp, integrating (\ref{wxt}) gives
 \be\label{wzt} -\big[\p_\bz A^{\0}_z +\p_zA^{\0}_\bz\big]_{v=\infty}=\int_{-\infty}^\infty dv\Big( [A^{\0}_z,\overleftrightarrow {\p_v}A^{\0}_\bz]+\gamma_{z\bz} g_{\rm YM}^2j^{\0 M}_v\Big).  \ee  
  Defining  
\be \label{dsf} J_z(z,\bz)=-{4\pi\over  g_{\rm YM}^2}A^\0_z(\infty, z,\bz), \ee
adding $F^\0_{z\bz}|_{v=\infty}=0$ to the left hand side of (\ref{wzt}) and then subtracting $[A^\0_z,A^\0_\bz]_{v=\infty}$ from both sides gives 
 \be\label{wbxt} \p_\bz J_z =2\pi \int_{-\infty}^\infty dv\Big(\gamma_{z\bz}j^{0 E}_v+{2 \over  g_{\rm YM}^2} [A^{\0}_\bz, {\p_v}A^{\0}_z]\Big)\equiv 2\pi q_{z\bz}.  \ee 
 This implies 
\be J_z(z)=\int d^2w {q_{w\bw} (w,\bw)\over (z-w)} .\ee
As in the abelian case the gauge variation of the curent 
 \be \delta_\e J_z=-{4\pi \over g_{YM}^2}\p_z\e+[J_z,\e]\ee
hints at a level proportional to 
${4\pi \over  g_{\rm YM}^2}$.
The Ward identity (\ref{xzc}) becomes 
\be \< V_1(v_1,z_1,\bz_1)...V_n(v_n, z_n,\bz_n)J^a_z(z,\bz)....\> =\sum_{k=1}^n{T^a_k \over z-z_k} \< V_1(v_1,z_1,\bz_1)...V_n(v_n, z_n,\bz_n)....\> \ee 
where we display  the adjoint index $a$ and $T^a_k$ is a generator in the representation of and acts on the operator insertion at $z_k$.  This is the soft gluon theorem transformed from momentum space to spherical  coordinates on \ip. 

\section{Stringy realization?}
 
 Our arguments so far have generally applied to any 4D gauge theory with massless charges. However the reader familiar with string theory may have found some features familiar from the study of string scattering amplitudes. The role of the conformal sphere at \ip\ resembles that of the string worldsheet: see for example \cite{gm,Bengtsson:2012dw}. Perhaps this connection can be made precise: we give here a few speculative comments. 
 
 Our analysis applies in particular to any perturbative closed string compactification to four dimensions with a non-abelian gauge symmetry $G$.
For such theories the string world sheet has a level $k_{ws}$ $G$-current algebra with an OPE  
\be \label{wsj} j^a(\rz)j^b(\rw)\sim {k_{ws} \delta^{ab}\over (\rz-\rw)^2}+ {f^{abc}j^c(\rw)\over (\rz-\rw)^2}.\ee
We use roman case to distinguish the world sheet coordinate $\rz$ from the spacetime coordinate $z$. 
The four dimensional space-time string coupling $g_{\rm S}$,  and world sheet level $k_{\rm ws}$ are related by \cite{Polchinski:1998rr}
\be \label{kt}{k_{\rm ws}\over g_{\rm S}^2} {V_6 \over 4(2\pi)^6}  ={4\pi \over  g_{\rm YM}^2},\ee
where $V_6$ is the compactification volume in string units.  On the other hand this paper raised the possibility of a relation of the form 
\be \label{kst} k_{\rm spacetime}\sim{4\pi \over  g_{\rm YM}^2}= {k_{\rm ws}\over g_{\rm S}^2} {V_6 \over 4(2\pi)^6}    .\ee

The world sheet current algebra (\ref{wsj}) generated by $j^a_\rz(\rz)$ is of course not manifestly the same as the spacetime current algebra generated by $J^a_z(z)$.  Nevertheless the action of our Kac-Moody on the conformal sphere at \ip\ resembles the action of the worldsheet Kac-Moody on the string worldsheet. It is often the case in string theory that worldsheet symmetries lift to space-time symmetries. Consider for example strings on $AdS_3\times S^3 \times M^4$ with NS fluxes. The string world sheet has a 
level $k_{ws}$ $SU(2)_L\times SU(2)_R$ current algebra.   As shown in \cite{Giveon:1998ns}, this lifts to a $spacetime$ current algebra in the boundary 
CFT$_2$ with level  
\be k_{\rm spacetime}= {k_{\rm ws}^2\over g_{\rm S}^2} {V_4  \over (2\pi)^4} ,\ee
where $V_4$ is the string-unit volume of $M_4$. This formula bears some similarity to (\ref{kst}), which one hopes might be derived by similar methods.

\section*{Acknowledgements}
I would like to thank J. Bourjailly, M. Guica, P. Mitra, G. S. Ng and W. Song  for stimulating conversations. I am particularly grateful to J. Maldacena for many discussions and for sharing the significant unpublished work \cite{juan}. 
This work is partly supported by DOE grant DE-FG02-91ER40654.

\end{document}